# Consistent description of the metallic phase of overdoped cuprate superconductors as an anisotropic marginal Fermi liquid

J. Kokalj[*] and Ross H. McKenzie

*School of Mathematics and Physics, University of Queensland, Brisbane, 4072 Queensland, Australia*

(Dated: July 19, 2011)

We consider a model self-energy consisting of an isotropic Fermi liquid term and a Marginal Fermi liquid term which is anisotropic over the Fermi surface, vanishing in the same directions as the superconducting gap and the pseudogap. This model self-energy gives a consistent description of experimental results from Angle-Dependent Magneto-Resistance (ADMR), specific heat, de Haas-van Alphen, and measurements of the quasi-particle dispersion near the Fermi surface from photoemission. In particular, we reconcile the strongly doping dependent anomalous scattering rate observed in ADMR with the almost doping independent specific heat.

A key to understanding high-$T_c$ superconductivity may be the anomalous properties of the metallic phase, which are quite distinct from those found in conventional Fermi liquids such as elemental metals. Many properties, such as the pseudogap, are strongly dependent on doping and on the position on the Fermi surface. One can attempt to describe the crucial effect of the strong electron-electron interactions by a frequency and momentum dependent electronic self-energy. Angle-Resolved Photoemission Spectroscopy (ARPES) [1] has been used to deduce various forms for the self-energy [2–5] including that of the marginal Fermi liquid phenomenology [6]. However, this approach implicitly assumes the existence of quasi-particles and the associated analytic structure of the one electron Green's function, which Anderson has contested and proposed an alternative "Hidden Fermi" liquid (HFL) theory [7].

In this Letter, we consider a model self-energy motivated by Angle-Dependent Magneto-Resistance (ADMR) experiments [8–10] and consisting of two terms with distinctly different dependencies on frequency, momentum, and temperature. The first term is that of a Fermi liquid (FL) and is isotropic on the Fermi surface. The second term, which we denote as an anisotropic marginal Fermi liquid (AMFL) has the same frequency and temperature dependence as that of a marginal Fermi liquid, is anisotropic over the Fermi surface, and vanishes in the same directions as the superconducting gap and the pseudogap observed in underdoped cuprates. We present a parametrization of this model self-energy which gives a consistent quantitative description of a wide range of experimental results on overdoped Tl2201 materials, including ADMR, specific heat [11], de Haas-van Alphen [12, 13] and the quasi-particle dispersion near the Fermi surface measured by ARPES [14, 15]. In particular, we give a consistent description of the *strongly doping dependent anisotropic scattering* [9] and the almost *doping independent specific heat* [11]. This is possible because although the scattering can be dominated by the AMFL term the quasi-particle renormalization is dominated by the FL term. We compare our parametrization of the self-energy with the results of different microscopic theories [16–19] based on Hubbard and $t-J$ models. In particular, we show that predictions of Hidden Fermi liquid theory for the temperature and doping dependence of the scattering rate and the magnitude of the specific heat [17, 20] are inconsistent with experiment.

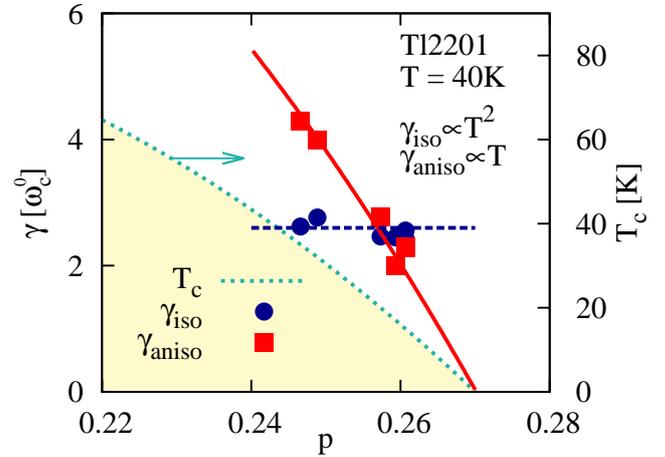

Figure 1. (color online) Strong doping dependence of the normal state anisotropic scattering rate in an overdoped cuprate superconductor. The scattering rate and its anisotropy are those deduced from the angular dependent magnetoresistance (ADMR) for Tl2201 at 40 K [9]. The isotropic scattering rate $\gamma_{\rm iso}$ (points and dashed line), which is constant on the Fermi surface, shows negligible doping dependence. In contrast, the anisotropic scattering rate $\gamma_{\rm aniso}$ (squares and full line), which is maximal in the antinodal direction and zero in the nodal direction, depends strongly on doping. The doping dependence of $\gamma_{\rm aniso}$ follows the superconducting transition temperature $T_c$ (dotted line) in this overdoped regime [9, 10]. The scattering rates are shown in units of $\omega_c^0 \simeq 1$ meV, the cyclotron frequency at the magnetic field at which the measurements were made.

*Is there a consistent phenomenology of the experiments?* For overdoped materials ADMR provides a complementary probe to ARPES, measuring the Fermi surface (FS) and the quasi-particle (QP) lifetime [8–10] at different points of the FS. Two scattering channels are observed, one has a quadratic temperature ($T$) dependence and is approximately constant with doping, while the second is approximately linear in $T$, anisotropic over the FS [8], and strongly increases with decreasing doping, as optimal doping ($p \simeq 0.16$) is approached from the overdoped regime (see Fig. 1) [9, 10]. Information on the self-energy is also provided through the renormalization of quasi-particle energies suggested by specific heat $C_V$ measurements [11], ARPES determination of Fermi velocity [21], de Haas-van Alphen (dHvA) measurements of the renormalized cyclotron mass [12, 13] and the optical effective mass determined from the Drude weight in the frequency dependent conductivity [22]. All of these suggest a weak doping dependence of the real part of the self-energy, in contrast to the strongly increasing anisotropic scattering rate $\gamma_{\text{aniso}}$ with decreasing doping shown in Fig. 1. This raises a question about consistency because the quasi-particle renormalization and scattering rate are not independent of one another, being related to the real and imaginary part of the self-energy, respectively. The two parts are related via the Kramers-Kronig relation. Indeed, this relationship is the origin of the unified picture of the Kadowaki-Woods ratio in Fermi liquids [23].

*Model self-energy.* Following the temperature dependence and anisotropy of the scattering rate determined by ADMR we consider a self-energy consisting of FL and AMFL contributions

$$\Sigma(\mathbf{k},\omega) = \Sigma_{\text{FL}}(\omega) + \Sigma_{\text{AMFL}}(\mathbf{k},\omega). \quad (1)$$

The detailed functional form is given in the Supplementary material. As suggested by the isotropic and $\propto T^2$ scattering rate in ADMR [8, 9] we take the FL self-energy isotropic. The AMFL part of the self-energy depends on $\phi$ [4] (azimuthal angle of the Fermi wave vector $\mathbf{k}_F$ on a 2D FS) and we assume that it is responsible for the anisotropic and $T$-linear part of the scattering deduced from ADMR [8].

Renormalization of the bare-band mass $m_b$ is determined by $1 - \partial\Sigma'(\phi,\omega)/\partial\omega|_{\omega=0}$, which for our model self-energy gives [24],

$$\frac{m^*(\phi)}{m_b} = Z(\phi)^{-1} = 1 + \frac{4}{\pi}\frac{s}{\omega^*_{\text{FL}}} + \lambda(\phi)\ln(\frac{\omega^*_{\text{AMFL}}}{\pi T}), \quad (2)$$

where $Z(\phi)$ is the QP weight at angle $\phi$. $s$ parametrizes the strength of the electron-electron scattering associated with the FL term, $\omega^*_{\text{FL}}$ is the FL high frequency cutoff, $\lambda(\phi)$ is a $\phi$ dependent dimensionless AMFL coupling constant, and $\omega^*_{\text{AMFL}}$ is the AMFL high frequency cutoff. We use units $\hbar = k_B = 1$. In general there is also a contribution to the renormalization from $\partial\Sigma/\partial k_\perp$ where $k_\perp$ is a momentum perpendicular to the FS. We assume this contribution is negligible [24].

*Parametrization of model self-energy.* In the Supplement we estimate the parameters in Eq. (2) from the ADMR results and show that $s/\omega^*_{\text{FL}} > \lambda(\phi)$ and that $\lambda(\phi)$ vanishes in the nodal direction. Together this leads to a relatively small effect of the AMFL part of the self-energy on the mass renormalization. Briefly, the isotropic $T^2$ term gives $s/\omega^{*2}_{\text{FL}} \simeq 9.2\,(\text{eV})^{-1}$. The term linear in $T$ gives the strength of the AMFL self-energy and its $\phi$ dependence [9],

$$\lambda(\phi) = 1.6\cos^2(2\phi)T_c(p)/T_c^{\max}, \quad (3)$$

where the doping dependence is encoded via the relation between $T_c$ and $p$ [24]. This expression for $\lambda(\phi)$ explicitly takes into account that the AMFL self-energy is largest in the antinodal direction, zero in the nodal direction [8, 10] and that it scales with $T_c$ in the highly overdoped regime [9]. The Fermi liquid cutoff $\omega^*_{\text{FL}} \simeq 0.23$ eV is estimated from measurements of $C_V$ in the strongly overdoped regime with $T_c = 0$ which give $m^*/m_e = 4.8 \pm 0.8$ [11]. Estimating the AMFL cutoff $\omega^*_{\text{AMFL}}$ is discussed below.

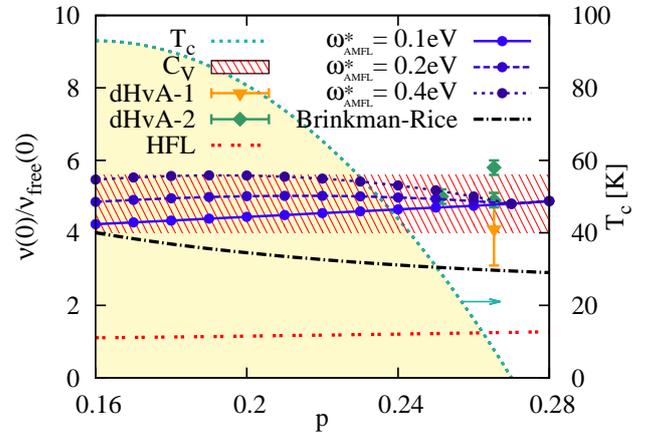

Figure 2. (color online) Renormalized density of states $\nu(0)/\nu_{\text{free}}(0)$ as a function of doping for the anisotropic marginal Fermi liquid model with various values of the cutoff frequency $\omega^*_{\text{AMFL}}$. The shaded area shows the range of measured values from $C_V$ [11]. Triangles (dHvA-1) and diamonds (dHvA-2) with error-bars show the effective masses deduced from de Haas-van Alphen effect in Ref. [12] and [13], respectively. Points with different lines shows our estimate deduced from the observed ADMR scattering rate (see Fig. 1) together with Eq. (2) for various $\omega^*_{\text{AMFL}}$, $T = 120$ K, and $\omega^*_{\text{FL}} = 0.23$ eV. Our estimates are within the measured uncertainties of $C_V$, which shows that strongly doping dependent ADMR scattering rates and doping independent $C_V$ can be consistently described. The Brinkman and Rice result [25] is shown with a dash-dotted line and the hidden FL result [20] is shown with a double-dotted line. The density of states is normalised to that for free electrons in two dimensions $\nu_{\text{free}}(0) = 4\pi m_e$.

*Renormalization factor.* Fig. 2 shows the calculated density of states at the Fermi energy $\nu(0)$ as a function of doping for various $\omega^*_{\text{AMFL}}$ together with values deduced from specific heat and de Haas-van Alphen experiments. For the calculation of $\nu(0)$ one evaluates the band mass $m_b$ (or strictly the band density of states at the Fermi energy) from the bare-band dispersion $\epsilon^0_{\mathbf{k}}$, which is approximated with a tight-binding model fit to the LDA bands (Eq. (S7) in Supplemental material [24]). It is evident from Fig. 2, that although the scattering rate of QP in the antinodal direction (or AMFL part of self-energy) strongly increases with decreasing doping, the density of states (and $C_V$) stays rather constant and is only mildly affected by the AMFL self-energy for all $\omega^*_{\text{AMFL}} \lesssim 0.4$ eV. This is because $m^*/m_b \sim 1 + 4s/\pi\omega^*_{\text{FL}} + \lambda(0)$ for $T \sim 120$ K. Subtleties associated with the relationship between the effective mass and $C_V$ for the AMFL are discussed in the Supplement [24].

Hence, it is possible for the model self-energy to give a con-sistent description of the complete doping and temperature dependence of both ADMR [8–10] and $C_V$ measurements [11] with $\omega_{FL}^* \simeq 0.23$ eV. The results are within measured uncertainty for any $\omega_{AMFL}^* \lesssim 0.5$ eV which is comparable to previous estimates from ARPES, $\sim 0.2$ eV - 0.4 eV [4], $\sim 0.1$ eV [3], and $\sim 0.4$ eV - 0.5 eV within the isotropic MFL phenomenology [5, 26]. From the above analysis it is evident that not only small $\omega$ properties are relevant, but also the high-energy cutoffs. These may be reflected as kinks or waterfalls in the QP dispersion [5, 27, 28].

*Renormalized QP dispersion.* Our renormalized dispersion is also in good agreement with the ARPES QP dispersion [14, 15] near $(\pi, 0)$ (see Fig. 3a), if a small correction of fixing the non-interacting FS to the one measured in ARPES [14] is taken into account by applying a bare-band shift of $(0.17\cos(4\phi) - 0.1\cos(8\phi))$ eV. The agreement in the nodal direction near the FS is satisfactory, but near the $(0, 0)$ point we observe the waterfall due to the sharp cutoff at $\omega_{FL}^*$ (Fig. 3b). The waterfall arises due to $\partial \Sigma'/\partial \omega$ becoming $\gtrsim 1$ in the vicinity of a high frequency cutoff $\omega_{FL}^*$ (for example see Fig. 1 in Ref. [5]). This results in a sharp drop of the QP dispersion and in a broader spectra at $\omega \sim \omega_{FL}^*$. The discrepancy at the band bottom may come from difficulties of determining the dispersion from a very broad ARPES spectra or may be an artifact of our approximation for self-energy [24].

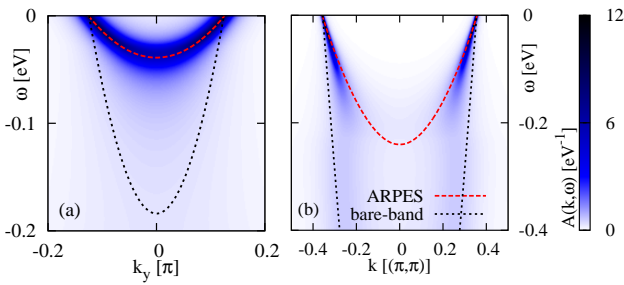

Figure 3. (color online) Comparison of AMFL spectral function with the quasi-particle dispersion measured by ARPES. Spectral functions calculated with our self-energy ($T = 10$ K, $T_c = 30$ K, $\omega_{AMFL}^* = 0.2$ eV, $\omega_{FL}^* = 0.23$ eV) are shown with density plots near the van Hove singularity for $\mathbf{k} = (\pi, k_y)$ (a) and near the band bottom in the nodal direction for $\mathbf{k} = (k, k)$ (b). The agreement with the measured ARPES QP dispersion [14] (dashed line) is very good near van Hove singularity and at the FS in the nodal direction (b). Some discrepancy is found at the band bottom, where we observe a waterfall originating from $\omega_{FL}^*$. The bare-band dispersion is shown with the dotted line.

Our estimate of $\lambda$ for the overdoped regime is in satisfactory agreement with the values estimated from ARPES, although some differences are still present [24]. The ARPES estimate for the QP lifetime on the FS [14] in Tl2201 is an order of magnitude larger than that from ADMR, and has the opposite angular dependence. This would imply that the renormalization of the effective mass and $C_V$ would be one order of magnitude larger, unless the scattering is elastic or some other effects, e.g., surface reconstruction, additionally broaden the ARPES spectra.

To partially conclude, our model self-energy is capable of describing a range of experimental results, including the strongly doping dependent ADMR scattering rate and almost doping independent $C_V$. Earlier it has been shown that the scattering rate deduced from ADMR can describe the temperature dependence of the intralayer resistivity and Hall coefficient [8, 9]. Future studies should examine whether for overdoped cuprates this model self-energy can describe the optical conductivity, asymmetry of tunneling spectra, and ARPES energy distribution curves, particularly since these have been invoked as evidence for the Hidden Fermi liquid theory [7].

*Microscopic theories.* We now turn to possible microscopic explanations of the self-energy and its dependence on $\phi$, $T$ and $p$ within the framework of strongly correlated electron lattice models. Fig. 3 illustrates the large difference between the bare-band dispersion and that observed by ARPES. They differ by a factor of about 4 due the QP renormalization from strong electronic correlations. It is a challenge for microscopic theory to explain these large renormalizations in the overdoped region. Fig. 2 illustrates how the renormalization is larger than the value $(1 + p)/(2p)$ predicted by the Brinkman-Rice theory [25]. However, it neglects the effect of the antiferromagnetic exchange interaction $J$. For small dopings this leads to an effective hopping of order $J + pt$ [18, 29, 30], and so this significantly reduces the QP renormalization compared to the Brinkman-Rice picture (meaning the discrepancy shown in Fig. 2 will be even greater). In the cuprates it is estimated that $J/t \simeq 0.3$ [31] and so this can explain the weak doping dependence of the renormalization, but not its large magnitude.

We now compare the scattering rate with microscopic theories. First, a weak coupling treatment of the Hubbard model produces an anisotropic scattering rate of similar frequency and angular dependence. The MFL component arises from a nesting of the Fermi surface in the anti-nodal regions [32] or from proximity to a van Hove singularity [32, 33]. However, for the later case the resulting scattering rate would have opposite doping dependence and would appear only at higher T than experimentaly observed for Tl2201. Furthermore, a functional renormalization group treatment of the Hubbard model shows scattering rates in qualitative agreement with ADMR [16]. However, we show in the Supplement that the calculated anisotropic scattering rate is an order of magnitude smaller than in experiment [24]. Other microscopic model calculations also give a scattering rate similar to our model form, and a quantitative comparison with our self-energy may rule out some. Candidates include cluster dynamical mean-field theory [34], a large-$N$ expansion treatment of the $t$-$J$ model [18], the quantum critical regime near a $d$ wave Pomeranchuk instability [19], and d-wave superconducting fluctuations [35].

*Hidden Fermi Liquid (HFL) theory.* Anderson has argued that the overdoped cuprates can be described in terms of a Gutzwiller projected Fermi liquid which exhibits power law singularities related to the X-ray edge problem [7]. Casey and Anderson calculated the scattering rate (see Eq. S10 in Supplement [24]) and compared it to ADMR data [17]. How-ever, the scattering rate in HFL [17] has a linear $T$ dependence only for $T \gtrsim W_{HFL}/2 \sim 400$ K [24], in strong contrast to the ADMR measurements [8], where the $T$ linear term is observed even for $T < 60$ K (see Fig. 4 and Fig. S1 in Supplement [24]). $W_{HFL}$ is a HFL bandwidth [24]. Furthermore, they argue that the anisotropic scattering rate deduced from ADMR emerges solely as a consequence of anisotropy of the Fermi momentum and of the Fermi velocity on the FS [17, 20]. To obtain anisotropies comparable to ADMR Casey and Anderson require that $v_{\mathrm{F}}(0)/v_{\mathrm{F}}(\pi/4) \simeq 0.5$. We note that the anisotropy in $v_{\mathrm{F}}(\phi)$ found in LDA calculations is smaller ($v_{\mathrm{F}}(0)/v_{\mathrm{F}}(\pi/4) \simeq 0.8$ [14, 36]). In addition the LDA calculations show that this ratio increases with increasing doping due to approaching the van Hove singularity [36]. Hence, HFL theory predicts a small increase in the ratio of anisotropic to isotropic scattering with increasing doping (this effect is not taken into account in Fig. 4); the opposite trend is observed with ADMR [9] (see Fig. 4 and S2). Furthermore, HFL predicts $C_V \propto T/\epsilon_F^0$ with no renormalization effects [20], which is significantly smaller than experimental results for Tl2201 (see Fig. 2) [37].

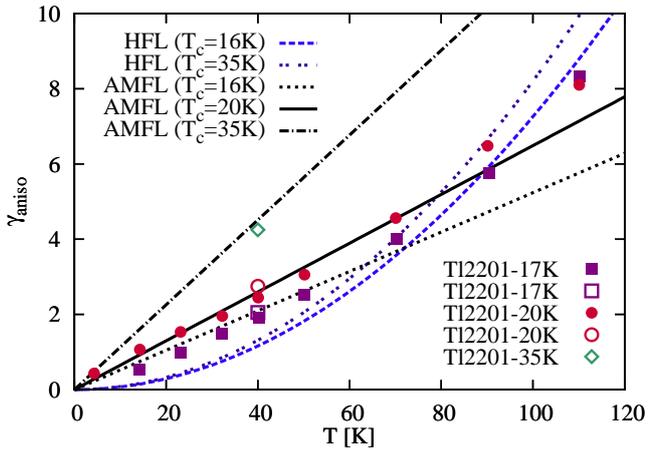

Figure 4. (color online) Hidden Fermi Liquid (HFL) theory cannot describe the temperature and doping dependence of the anisotropic scattering, $\gamma_{\mathrm{aniso}}$. HFL predictions [17] for $\gamma_{\mathrm{aniso}}$ are compared with values deduced from ADMR data [9, 10] and to our anisotropic marginal Fermi liquid (AMFL) parametrisation of the ADMR results. HFL predictions [17] for dopings corresponding to $T_c = 16$ and 35 K are shown with dashed and double-dotted lines, respectively. HFL theory does not show any significant linear $T$ dependence at low temperatures, unlike the measured data. (Full symbols are taken from Ref. [10], empty symbols are from Ref. [9]). Furthermore, HFL shows negligible doping dependence, unlike the significant doping dependence seen in the experimental data. Our AMFL parametrisation of the data is shown with dotted, full and dash-dotted lines. HFL results are obtained with $v_{\mathrm{F}}(0)/v_{\mathrm{F}}(\pi/4) = 0.5$ (as in [17]), which is kept constant with $p$.

In conclusion, we have shown that it is possible to give a consistent description of a wide range of experimental results for overdoped cuprates in terms of a model self-energy which contains an isotropic Fermi liquid contribution and an anisotropic marginal Fermi liquid contribution. The former is doping independent and the latter increases significantly with decreasing doping. This model self-energy is quantitatively inconsistent with some microscopic model calculations and provides a explicit form against which other calculations can be compared. The two distinct terms in the self-energy may have two distinct physical origins. The isotropic Fermi liquid terms arises largely from local physics. The large on site Coulomb repulsion $U$ reduces intersite hopping and leads to Fermi liquid scattering of quasi-particles. Qualitatively, this can be captured in a Brinkman-Rice picture and by Dynamical Mean-Field Theory (DMFT). In contrast, the anisotropic marginal Fermi liquid term arises from non-local physics, and its physical origin is unclear. The relative importance of different types of fluctuations (antiferromagnetic, superconducting, or d-density wave), Fermi surface nesting and proximity to a quantum critical point is unclear.

This work shows that the overdoped curates are not simple Fermi liquids as has often been claimed. Instead they exhibit remnants of some the same physics present in the optimally doped materials (marginal Fermi liquid behaviour) and the underdoped materials (cold spots and well-defined quasi-particles at the same Fermi surface points as the nodes in the superconducting gap and pseudogap). Thus, it seems that the challenge of finding a successful microscopic theoretical description of the metallic phase of the cuprates is now extended to the overdoped regime.

This work was stimulated by discussions with N.E. Hussey. We also acknowledge discussions with A. Carrington, M. Kennett, and B.J. Powell. We thank P.W. Anderson, A. Jacko, and Y. Zhang for comments on the manuscript. Financial support was received from an Australian Research Council Discovery Project grant (DP1094395).

* j.kokalj@uq.edu.au; On leave from J. Stefan Institute, Ljubljana, Slovenia
[1] A. Damascelli, Z. Hussain, Z.X. Shen, Rev. Mod. Phys. **75**, 473 (2003).
[2] T. Valla, A.V. Fedorov, P.D. Johnson, B.O. Wells et al., Science **285**, 2110 (1999).
[3] A.A. Kordyuk, S.V. Borisenko, A. Koitzsch, J. Fink et al., Phys. Rev. Lett. **92**, 257006 (2004).
[4] J. Chang, M. Shi, S. Pailhés, M. Månsson et al., Phys. Rev. B **78**, 205103 (2008).
[5] L. Zhu, V. Aji, A. Shekhter, C.M. Varma, Phys. Rev. Lett. **100**, 057001 (2008).
[6] C.M. Varma, Int. J. Mod. Phys. B **3**, 2083 (1989).
[7] P.W. Anderson, Nature Phys. **2**, 626 (2006).
[8] M. Abdel-Jawad, M.P. Kennett, L. Balicas, A. Carrington et al., Nature Phys. **2**, 821 (2006).
[9] M. Abdel-Jawad, J.G. Analytis, L. Balicas, A. Carrington et al., Phys. Rev. Lett. **99**, 107002 (2007).
[10] M.M.J. French, J.G. Analytis, A. Carrington, L. Balicas et al., New J. Phys. **11**, 055057 (2009).
[11] J.M. Wade, J.W. Loram, K.A. Mirza, J.R. Cooper et al., J. Supercond. **7**, 261 (1994); J.W. Loram, K.A. Mirza, J.M. Wade,

J.R. Cooper et al., Physica C **235-240**, 134 (1994).
[12] B. Vignolle, A. Carrington, R.A. Cooper, M.M.J. French et al., Nature Phys. **455**, 952 (2008).
[13] A.F. Bangura, P.M.C. Rourke, T.M. Benseman, M. Matusiak et al., Phys. Rev. B **82**, 140501 (2010).
[14] M. Platé, J.D.F. Mottershead, I.S. Elfimov, D.C. Peets et al., Phys. Rev. Lett. **95**, 077001 (2005).
[15] D.C. Peets, J.D.F. Mottershead, B. Wu, I.S. Elfimov et al., New J. Phys. **9**, 28 (2007).
[16] M. Ossadnik, C. Honerkamp, T.M. Rice, M. Sigrist, Phys. Rev. Lett. **101**, 256405 (2008).
[17] P.A. Casey, P.W. Anderson, Phys. Rev. Lett. **106**, 097002 (2011).
[18] G. Buzon, A. Greco, Phys. Rev. B **82**, 054526 (2010).
[19] L. Dell'Anna, W. Metzner, Phys. Rev. Lett. **98**, 136402 (2007).
[20] P.A. Casey (Ph.D. thesis, Princeton Physics, 2010), p. 79.
[21] X.J. Zhou, T. Yoshida, A. Lanzara, P.V. Bogdanov et al., Nature Phys. **423**, 398 (2003).
[22] W.J. Padilla, Y.S. Lee, M. Dumm, G. Blumberg et al., Phys. Rev. B **72**, 060511 (2005).
[23] A.C. Jacko, J.O. Fjaerestad, B.J. Powell, Nature Phys. **5**, 422 (2009).
[24] See the Supplemental material.
[25] W.F. Brinkman, T.M. Rice, Phys. Rev. B **2**, 4302 (1970).
[26] J.M. Bok, J.H. Yun, H.Y. Choi, W. Zhang et al., Phys. Rev. B **81**, 174516 (2010).
[27] K. Byczuk, M. Kollar, K. Held, Y.F. Yang et al., Nature Phys. **3**, 168 (2007).
[28] M.M. Zemljič, P. Prelovšek, T. Tohyama, Phys. Rev. Lett. **100**, 036402 (2008).
[29] G. Kotliar, J. Liu, Phys. Rev. B **38**, 5142 (1988).
[30] G. Martinez, P. Horsch, Phys. Rev. B **44**, 317 (1991).
[31] E. Dagotto, Rev. Mod. Phys. **66**, 763 (1994).
[32] R. Roldán, M.P. López-Sancho, F. Guinea, S.W. Tsai, Phys. Rev. B **74**, 235109 (2006).
[33] G. Kastrinakis, Phys. Rev. B **71**, 014520 (2005).
[34] M. Civelli, M. Capone, S.S. Kancharla, O. Parcollet et al., Phys. Rev. Lett. **95**, 106402 (2005).
[35] L.B. Ioffe, A.J. Millis, Phys. Rev. B **58**, 11631 (1998).
[36] J.G. Analytis, M. Abdel-Jawad, L. Balicas, M.M.J. French et al., Phys. Rev. B **76**, 104523 (2007).
[37] The HFL result corresponds to the bare-band density of states given, e.g., with the LDA in Ref. [15] or in Ref. [38], where slightly larger values are obtained.
[38] P.M.C. Rourke, A.F. Bangura, T.M. Benseman, M. Matusiak et al., New J. Phys. **12**, 105009 (2010).

# Supplemental material for "Consistent description of the metallic phase of overdoped cuprate superconductors as an anisotropic marginal Fermi liquid"

J. Kokalj and Ross H. McKenzie

*School of Mathematics and Physics, University of Queensland, Brisbane, 4072 Queensland, Australia*

(Dated: July 19, 2011)

## MODEL SELF-ENERGY

We take the FL part of the self-energy isotropic and parametrize its imaginary part with [1]

$$\Sigma''_{\text{FL}}(\omega) = \begin{cases} -\frac{1}{2\tau_0} - s\frac{\omega^2+\pi^2 T^2}{\omega^{*2}_{\text{FL}}} & \text{for } \frac{\omega^2+\pi^2 T^2}{\omega^{*2}_{\text{FL}}} \leq 1, \\ \left[-\frac{1}{2\tau_0} - s\right] F\left(\frac{\omega^2+\pi^2 T^2}{\omega^{*2}_{\text{FL}}}\right) & \text{for } \frac{\omega^2+\pi^2 T^2}{\omega^{*2}_{\text{FL}}} > 1. \end{cases} \quad (S1)$$

Here $1/\tau_0$ accounts for impurity scattering, $s$ parametrizes the strength of electron-electron scattering, $\omega^*_{\text{FL}}$ is the high frequency cutoff and $F(y)$ is monotonically decreasing function with $F(1) = 1$. We use units $\hbar = k_B = 1$. The real part of $\Sigma_{\text{FL}}(\omega)$ is obtained from Eq. (S1) with a Kramers-Kronig transformation, which for the case of slowly decreasing $F(y)$ and for $T, \omega \ll \omega^*_{\text{FL}}$ yields [2]

$$\Sigma'_{\text{FL}}(\omega) = -\frac{4}{\pi} s \frac{\omega}{\omega^*_{\text{FL}}}. \quad (S2)$$

We parametrize the imaginary part of the AMFL self-energy with [3–5]

$$\Sigma''_{\text{AMFL}}(\phi,\omega) = \begin{cases} \lambda(\phi)(-\frac{\pi}{2}x) & \text{if } |\omega| \leq \omega^*_{\text{AMFL}}, \\ \lambda(\phi)(-\frac{\pi}{2}\omega^*_{\text{AMFL}}) & \text{if } |\omega| > \omega^*_{\text{AMFL}}, \end{cases} \quad (S3)$$

where $\lambda(\phi)$ is a $\phi$ dependent dimensionless coupling constant, $x = \max(|\omega|, \pi T)$, and $\omega^*_{\text{AMFL}}$ is the AMFL high frequency cutoff. The real part of the AMFL self-energy is obtained with the Kramers-Kronig transformation, which for $\omega, T < \omega^*_{\text{AMFL}}$ gives

$$\Sigma'_{\text{AMFL}}(\phi,\omega) = -\lambda(\phi)\omega \ln\left(\frac{\omega^*_{\text{AMFL}}}{x}\right). \quad (S4)$$

## PARAMETRIZATION OF MODEL SELF-ENERGY

We estimate the parameters in Eqs. (S1) and (S3) from the ADMR results and show that $s/\omega^*_{\text{FL}} > \lambda(\phi)$ and that $\lambda(\phi)$ vanishes in the nodal direction. Together this leads to a relatively small effect of the AMFL part of the self-energy on the mass renormalization (Eq. (2) in the main text). In order to parametrize the self-energy from the ADMR, which is predominantly measuring the lifetime $\tau(\phi)$ of QPs on the FS [6] (see Section *Connection of the ADMR Scattering Rate and the Self-energy*), we use the relation

$$\frac{1}{\tau(\phi)} = -2Z(\phi)\Sigma''(\phi,\omega=0), \quad (S5)$$

where we have explicitly denoted the $\phi$ dependence. To evaluate the imaginary part $\Sigma''(\phi,\omega=0)$ from Eq. (S5) the renormalization $Z(\phi)$ should be known in advance, but this can be avoided (see Eq. (S6) and the text above). The connection of the self-energy to the ADMR is given by $\Sigma''(\phi,\omega=0) \propto 1/(\omega_c(\phi)\tau(\phi))$, where the proportionality factor depends on bare-band Fermi velocity and Fermi wave vector, but is fairly constant with $\phi$ (Eq. (S6)).

The main $\phi$, $T$ and $p$ dependencies of the self-energy can be deduced from $1/\omega_c(\phi)\tau(\phi)$ (Eq. (S6)). Taking the measured temperature dependency of the QP lifetime [7, 8], $1/\omega_c(\phi)\tau(\phi) = a + bT^2 + c(\phi)T$, using the connection to the self-energy, Eq. (S6), and comparing it with the model self-energy, Eq. (1) in the main text, the following estimates are obtained. From the $T^2$ term and the parameter $b$ one can evaluate the strength of the FL self-energy part, $s/\omega^{*2}_{\text{FL}} \simeq 9.2(\text{eV})^{-1}$. The term linear in $T$ and $c(\phi)$ gives the strength of the AMFL self-energy and its $\phi$ dependence, Eq. (3) in the main text.

To calculate the density of QP states $\nu(0)$ at the Fermi energy and the mass renormalization two additional parameters are needed, the cutoffs $\omega^*_{\text{FL}}$ and $\omega^*_{\text{AMFL}}$. $\omega^*_{\text{FL}}$ influences the renormalization and the value of $\nu(0)$ quite strongly (Eq. (2) in the main text) and can be estimated in the strongly overdoped regime with $T_c = 0$ from the $C_V$ measurements. Using $m^*/m_e = 4.8 \pm 0.8$ [9] we estimate $\omega^*_{\text{FL}} \simeq 0.23$ eV, in good agreement with the relation [10] $\omega^*_{\text{FL}} \simeq 1.6 Z_{\text{FL}} t_1 \simeq 0.2$ eV, where $Z_{\text{FL}} \simeq m_b/m^* \simeq 0.3$ (using $m_b/m_e \simeq 1.5$ [11]).

The parameter $\omega^*_{\text{AMFL}}$ is in the argument of a logarithm in the renormalization Eq. (2) in the main text and therefore does not influence the renormalization considerably. Since it is observed that the QP density of states does not vary much as doping is decreased from an overdoped regime towards optimal doping [9], we can still try to set some limits on $\omega^*_{\text{AMFL}}$.

## MOMENTUM DEPENDENCE OF THE CUTOFF FREQUENCY $\omega^*_{\text{AMFL}}$

ARPES momentum distribution curves shown in Fig. 3 of Ref. [4] suggest that the high frequency cutoff may be momentum dependent, i.e., $\omega^*_{\text{AMFL}} = \omega^*_{\text{AMFL}}(\phi)$. Similarly, a $\phi$ dependent cutoff was observed by high-resolution laser ARPES for slightly underdoped cuprate (see Fig. 5 and 7 in Ref. [12]). In our analysis the $\phi$ dependency of $\omega^*_{\text{AMFL}}$ would be important only for $\phi$ for which the $\lambda(\phi)$ is large, i.e., in the antinodal direction. Our approximation of isotropic $\omega^*_{\text{AMFL}}$ can therefore be viewed as $\omega^*_{\text{AMFL}} = \omega^*_{\text{AMFL}}(0)$. On the other hand, different cutoffs in momentum distribution curves may be a consequence of a band dispersion as was shown in Ref. [13] (see Fig. 3), where an isotropic cutoff was used.

### $\partial\Sigma/\partial k_\perp$ CONTRIBUTION TO THE RENORMALIZATION

The contribution of $\partial\Sigma/\partial k_\perp$ to the renormalization is claimed to be negligible according to Ref. [14] and large in underdoped regime according to Ref. [15]. This contribution is finite for a self-energy, whose imaginary part is not an even function of $\omega$, since then $\partial\Sigma'(\omega=0)/\partial k_\perp \neq 0$. Strong $k_\perp$ dependency of the self-energy is also expected to change the FS from the non-interacting one considerably, but it was shown that the deviations are fairly small at least in the overdoped regime (see Fig. 9 in Ref. [11]). The FS does not show any significant change of squareness [16] with doping, which would signal the $k_\perp$ renormalization from AMFL part of the self-energy. Hence, we assume that $\partial\Sigma/\partial k_\perp$ contribution to the renormalization is negligible.

Within our model, there is also no change of the FS from the non-interacting one since $\Sigma'(\phi,\omega=0)=0$ for any $\phi$, which is a consequence of $\Sigma''(\phi,\omega)$ being an even function of $\omega$.

### CONNECTION OF THE ADMR SCATTERING RATE AND THE SELF-ENERGY

Eq. (3) in the main text raises the question whether the lifetime measured in ADMR is not rather the transport lifetime $1/\tau_{tr}$, which does not include the contribution of small angle scattering, while the QP lifetime $1/\tau$ does. More detailed analysis [6] shows that for $B$ perpendicular to the layers $1/\tau_{tr}$, while for $B$ parallel to the layers $1/\tau$, should be used in a Boltzmann equation. However, for scattering from atomic scale defects $1/\tau_{tr}$ and $1/\tau$ differ by at most a factor of order unity and for scattering potential with short-range interlayer correlations $1/\tau_{tr} \approx 1/\tau$ [6]. In this respect, Eq. (3) in the main text is a good approximation to the connection between the ADMR scattering rate and the self-energy. It is essentially the relaxation time approximation which is used in the analysis of ADMR data, and can be derived for a Boltzmann equation [6, 17].

To avoid the factor $Z(\phi)$ in calculation of $\Sigma''(\phi,\omega=0)$ from the ADMR scattering rate, one can use the observation that in ADMR the product $1/\omega_c(\phi)\tau(\phi)$ is actually measured [16] and that the cyclotron frequency $\omega_c(\phi) \propto v_F(\phi) \propto Z(\phi)v_F^0(\phi)$, where $\mathbf{v}_F(\phi)$ is the QP Fermi velocity and $\mathbf{v}_F^0(\phi)$ is a bare-band Fermi velocity. Using this together with $\omega_c(\phi) = eB\mathbf{v}_F(\phi)\cdot\mathbf{k}_F(\phi)/k_F^2(\phi)$ [18] and expressing the self-energy with the product $1/\omega_c(\phi)\tau(\phi)$ one obtains

$$\Sigma''(\phi,\omega=0) = -\frac{eB\mathbf{v}_F^0(\phi)\cdot\mathbf{k}_F(\phi)}{2k_F^2(\phi)}\left(\frac{1}{\omega_c(\phi)\tau(\phi)}\right), \quad \text{(S6)}$$

where the factor $Z(\phi)$ has canceled out. $e$ is an electron charge, $B$ is the magnetic field used in the ADMR measurements and $k_F(\phi)$ is a Fermi wave vector measured from $(\pi,\pi)$.

Comparison of the measured self-energy at $\omega=0$, Eq. (S6), and our model for the self-energy, Eq. (1) in main text, allows us to make quantitative estimates of parameters for the self-energy in the whole $\omega$ domain.

### BARE-BAND DISPERSION

The bare-band Fermi velocity $\mathbf{v}_F^0(\phi)$ is evaluated from the bare-band dispersion [19]

$$\begin{aligned}\epsilon_\mathbf{k}^0 = &\epsilon_0 - 2t_1(\cos k_x + \cos k_y) - 4t_2 \cos k_x \cos k_y \\ &-2t_3(\cos 2k_x + \cos 2k_y) \\ &-4t_4(\cos 2k_x \cos k_y + \cos k_x \cos 2k_y) \\ &-4t_5 \cos 2k_x \cos 2k_y \\ &-2t_6(\cos 3k_x + \cos 3k_y), \end{aligned} \quad \text{(S7)}$$

with parameters $\epsilon_0 = -1.598$, $t_1 = 0.438$, $t_2 = -0.150$, $t_3 = 0.084$, $t_4 = -0.013$, $t_5 = -0.020$, $t_6 = 0.029$, all expressed in eV. These parameter values are taken to obtain a good description of the LDA calculations presented in Fig. 7 in Ref. [19]. For different doping levels we assume a rigid band shift and apply the corresponding chemical potential. The Fermi wave vector $k_F(\phi) = k_{00}(p) - k_{40}\cos 4\phi$ [16] with $k_{00}(p)$ determined according to the Luttinger theorem $\pi(k_{00}(p))^2/(2\pi/a_0)^2 = (1+p)/2$, provided $k_{40} \ll k_{00}$ and as used in Ref. [16, 20]. $k_{40} = 0.034$ Å$^{-1}$ [8, 16] and lattice parameter $a_0 = 3.86$ Å. The measured $\mathbf{k}_F(\phi)$ is in satisfactory agreement with the FS obtained from $\epsilon_\mathbf{k}^0$ above. The prefactor $\mathbf{v}_F^0(\phi)\cdot\mathbf{k}_F(\phi)/k_F^2(\phi)$ in Eq. (S6), evaluated with the presented parameters and for the cases measured with ADMR [7, 8], shows less than 10% variation over the FS and can therefore be to a good approximation treated as a constant.

### CORRECTION TO THE SPECIFIC HEAT CAPACITY $C_V$ FOR AMFL

When calculating the density of states for AMFL and comparing it with the measured $C_V$, one should be cautious. It is not simply a matter of using the renormalized effective mass in a Fermi liquid expression for $C_V(T)$. The temperature dependence of the thermodynamic potential $\Omega(T)$ can be found from evaluating $\Omega(T)$ from the Luttinger-Ward expression in terms of the one-electron Green's function [21]. This free energy $\Omega(T)$ can be used to calculate the entropy and $C_V$ by differentiation with respect to $T$. This brings additional correction to the AMFL renormalization when dealing with $C_V$, since the real part of AMFL self-energy has a logarithmic $T$ dependence.

The isotropic marginal Fermi liquid expression for $C_V$ is

[22]

$$C_V \simeq \frac{2}{3}\pi^2 \nu_B(0)(1 + \lambda(\ln\left(\frac{\omega_{\text{MFL}}}{\pi T}\right) - 1))T, \quad \text{(S8)}$$

where $\nu_B(0)$ is a bare band density of states, $\lambda$ is the coupling constant and $\omega_{\text{MFL}}$ is a Marginal Fermi Liquid high frequency cutoff. The correction to the mass renormalization in Eq. (2) in the main text when dealing with $C_V$ is therefore to replace $\ln(\omega^*_{\text{AMFL}}/\pi T) \to (\ln(\omega^*_{\text{AMFL}}/\pi T) - 1)$. This correction is used in the calculation of $\nu(0)$, plotted in Figure 2 in the main text, which is mainly compared to $C_V$ measurements. However, the correction has no significant effect on the final results, particularly because the effective mass renormalisation is dominated by the isotropic Fermi liquid contribution.

## RELATION BETWEEN $T_c$ AND $p$

The relation between $T_c$ and $p$ used in Figs. 1 and 2 in the main text is the parabolic universal phenomenological relation [23],

$$\frac{T_c(p)}{T_c^{\max}} = 1 - 82.6(p - 0.16)^2 \quad \text{(S9)}$$

with $T_c^{\max} = 93$ K for Tl2201. This relation also gives the doping dependence of $\lambda(\phi)$ in the AMFL self-energy via $T_c$ in Eq. (3) in the main text. Including the observed deviation from the universal relation [24] would shift and stretch the values on the doping axis in Figs. 1 and 2 in the main text. For example, the value of $p$ at which $T_c$ becomes non-zero would be changed to 0.31, and the dHvA data in Fig. 2 in the main text for $T_c = 10$ and 26 K would have the doping values $p = 0.30$ and $0.27$, respectively. However, for the simplicity we use the usual phenomenological relation and note that the above changes in the relation between $T_c$ and $p$ would not change the overall results.

## COMPARISON WITH ARPES SPECTRA: WATERFALL AND $\omega^*_{\text{FL}}$ DISCUSSION

The waterfall would diminish from our calculations, if in Eq. (S1) for $\Sigma''_{\text{FL}}(\omega)$ a stronger decrease for high $\omega$ is used, which would result in smaller FL renormalization in Eq. (S2), but would in turn be compensated with larger $\omega^*_{\text{FL}}$. This would move the waterfall to higher frequencies or even diminish it for large enough $\omega^*_{\text{FL}}$. However, even for such $\omega^*_{\text{FL}}$, our QP dispersion at the band bottom would be $\sim 50$ meV lower than the one in Ref [25].

## COMPARISON WITH ARPES SPECTRA: DISCUSSION OF THE AMFL COUPLING CONSTANT $\lambda$

Our estimate of $\lambda$ for the overdoped regime is smaller than the values of $\lambda \sim 1 - 3$ estimated from ARPES for optimally doped LSCO [4]. However, taking into account that our $\lambda$, estimated from ADMR on a different material, is increasing with approaching optimal doping and reaches $\lambda \sim 1.6$ the agreement is satisfactory (see Eq. (3) in the main text). In addition, small $\lambda \sim 0.2$ has been reported for overdoped Bi(Pb)-2212 together with its increase with lowering doping [26], which is in good agreement with our estimate. However, ARPES estimates for $\lambda(\phi)$ [4, 26] do not completely vanish in nodal direction, but still shows significant increase towards antinodal direction [4].

On the other hand, the ARPES observation of increasing QP lifetime towards the antinodal direction [25] in overdoped Tl2201 could be the consequence of a SC phase [27], while in the normal state the QP lifetime could still be longer in the nodal direction.

## QUANTITATIVE COMPARISON WITH OSSADNIK ET AL. [28]

As discussed in the main text a functional renormalisation group treatment of the Hubbard model gives an anisotropic scattering rate which has qualitatively the same temperature and doping dependence as the experiments [28].

Our estimates from ADMR for the term linear in $T$ in the QP lifetime are $0.4T$, $2.5T$, $4T$, for dopings $p = 0.30, 0.22$, and $0.15$, respectively. These values should be compared with the values of $0.012T$, $0.09T$ and $0.2T$ obtained by Ossadnik et al. [28]

On the other hand, the prefactor in the $T^2$ term in the self-energy obtained in Ref. [28] is of the same order of magnitude as ours, provided a value is used for the nearest neighbour hopping $t_1 \simeq 0.44$ eV, consistent with LDA calculations [19], rather than the smaller one (0.18 eV) used in Ref. [28].

## HIDDEN FERMI LIQUID THEORY

Casey and Anderson argue that at a point on the Fermi surface and at temperature $T$ the scattering rate is [29]

$$\frac{1}{\omega_c \tau(\phi, T)} = \left[\frac{2\pi p' \hbar}{eB}\right] \left[\frac{k_F(\phi)}{v_F(\phi)}\right]$$
$$\times \frac{T^2}{T + 2\pi p' W_{HFL}^{(\pi,\pi)} \left[\frac{k_F(\phi)}{k_F(\pi/4)}\right] \left[\frac{v_F(\phi)}{v_F(\pi/4)}\right]} \quad \text{(S10)}$$

where the HFL bandwidth $W_{HFL} = \epsilon_F(2p/(1+p))^2$ and $p' = (1-p)^2/4$.

All the parameters we use in Figure 4 of the main text and in Figure S1 and S2 here are the same as those used by Casey and Anderson [29]. They compared the above equation with experimental data for a single doping and used the same vertical scale as in Figure S1 (top). There is then no clear discrepancy between the data and HFL theory. However, Figure 4 in the main text uses an expanded vertical scale and shows experimental data for a range of dopings. The experimental data is inconsistent with HFL theory.

In Figure S2 we explicitly show the measured doping dependence of anisotropic and isotropic scatterings and compare them with the HFL theory and our AMFL form. One can see that the HFL theory predicts a much weaker doping dependence of the anisotropic scattering than is observed, while it captures the observed negligible doping dependence of the isotropic scattering.

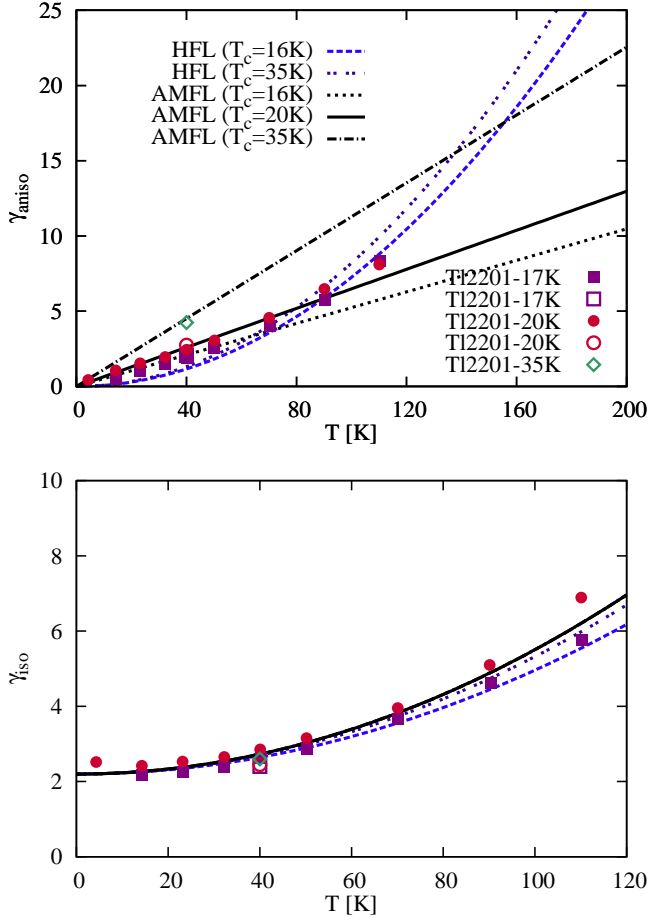

Figure S1. (color online) Top: Comparison of the measured anisotropic scattering rate $\gamma_{aniso}$ [8, 16] with HFL theory and our AMFL parametrization of ADMR results. Data are the same as in Fig. 4 in the main text but the vertical plot range is the same as Fig. 2 in Ref. [29]. Full squares and dashed line reproduce the plot in Fig. 2 in Ref. [29].
Bottom: Both HFL theory [29] and our AMFL form can successfully describe the measured isotropic scattering rate $\gamma_{iso}$ [8, 16]. Full squares and dashed line reproduce the plot in Fig. 2 in Ref. [29].

**FURTHER DISCUSSION**

We have shown that the renormalization coming from the AMFL part of the self-energy influences the $C_V$ only weakly and that the main source of renormalization comes from FL part of the self-energy. This sheds doubts on a purely AMFL self-energy as well as on the picture of increasing AMFL part and decreasing FL part of self-energy as optimal doping is approached from the overdoped side [4], since this should result in decrease of $C_V$.

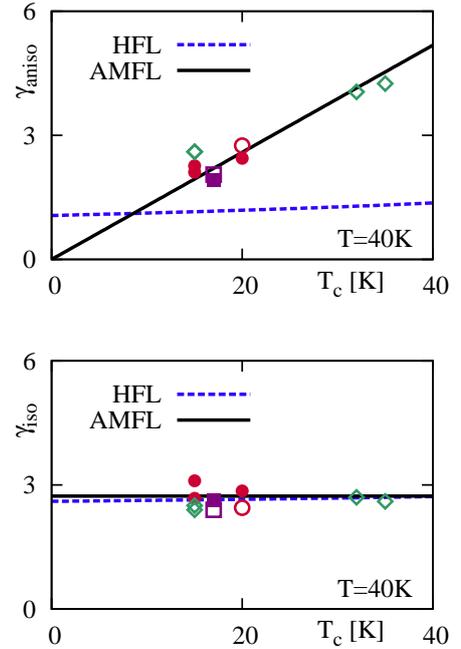

Figure S2. (color online) Doping dependence of $\gamma_{aniso}$ (top) and $\gamma_{iso}$ (bottom) at $T = 40$ K with the HFL theory prediction and our AMFL parametrization of ADMR results. Empty symbols are ADMR data from Ref. [16] and full symbols are ADMR data from Ref. [8].

However, the MFL renormalization can still be the only and sufficient source of the renormalization, if $\lambda$ is non-zero on the whole FS and larger $\omega_{MFL}^*$ are used, e.g., average $\lambda \sim 1.7$ and $\omega_{MFL}^* \sim 0.4$ eV used for optimally doped LSCO [13]. In the underdoped regime, $C_V$ shows a reduction [30], which is believed to be a consequence of the emerging pseudogap. In this Letter we are, however, using a QP picture, which becomes questionable as optimal doping is approached.

Eq. (2) in the main text shows that the AMFL renormalization logarithmically decreases with increasing temperature $T$, while the FL renormalization is fairly constant. Therefore the $T$ dependence of $C_V$ might prove useful for extracting the two contributions in the overdoped regime of high-$T_c$ superconductors, especially, if the AMFL renormalization dominates [13]. Experimental uncertainties and $T$ ranges seem to currently be insufficient for making such an analysis [30].

[1] K. Miyake, T. Matsuura, C. Varma, Solid State Comm. **71**, 1149 (1989).
[2] A.C. Jacko, J.O. Fjaerestad, B.J. Powell, Nature Phys. **5**, 422 (2009).
[3] C.M. Varma, P.B. Littlewood, S. Schmitt-Rink, E. Abrahams, A.E. Ruckenstein, Phys. Rev. Lett. **63**, 1996 (1989).

[4] J. Chang, M. Shi, S. Pailhés, M. Månsson, T. Claesson, O. Tjernberg, A. Bendounan, Y. Sassa, L. Patthey, N. Momono et al., Phys. Rev. B **78**, 205103 (2008).
[5] T. Valla, A.V. Fedorov, P.D. Johnson, B.O. Wells, S.L. Hulbert, Q. Li, G.D. Gu, N. Koshizuka, Science **285**, 2110 (1999).
[6] M.F. Smith, R.H. McKenzie, Phys. Rev. B **77**, 235123 (2008).
[7] M. Abdel-Jawad, M.P. Kennett, L. Balicas, A. Carrington, A.P. Mackenzie, R.H. McKenzie, N.E. Hussey, Nature Phys. **2**, 821 (2006).
[8] M.M.J. French, J.G. Analytis, A. Carrington, L. Balicas, N.E. Hussey, New J. Phys. **11**, 055057 (2009).
[9] J.M. Wade, J.W. Loram, K.A. Mirza, J.R. Cooper et al., J. Supercond. **7**, 261 (1994); J.W. Loram, K.A. Mirza, J.M. Wade, J.R. Cooper et al., Physica C **235-240**, 134 (1994).
[10] K. Byczuk, M. Kollar, K. Held, Y.F. Yang, I.A. Nekrasov, T. Pruschke, D. Vollhardt, Nature Phys. **3**, 168 (2007).
[11] P.M.C. Rourke, A.F. Bangura, T.M. Benseman, M. Matusiak, J.R. Cooper, A. Carrington, N.E. Hussey, New J. Phys. **12**, 105009 (2010).
[12] J.M. Bok, J.H. Yun, H.Y. Choi, W. Zhang, X.J. Zhou, C.M. Varma, Phys. Rev. B **81**, 174516 (2010).
[13] L. Zhu, V. Aji, A. Shekhter, C.M. Varma, Phys. Rev. Lett. **100**, 057001 (2008).
[14] R. Haslinger, A. Abanov, A. Chubukov, Europhys. Lett. **58**, 271 (2002).
[15] M. Randeria, A. Paramekanti, N. Trivedi, Phys. Rev. B **69**(14), 144509 (2004).
[16] M. Abdel-Jawad, J.G. Analytis, L. Balicas, A. Carrington, J.P.H. Charmant, M.M.J. French, N.E. Hussey, Phys. Rev. Lett. **99**, 107002 (2007).
[17] J. Rammer, H. Smith, Rev. Mod. Phys. **58**, 323 (1986).
[18] M.P. Kennett, R.H. McKenzie, Phys. Rev. B **76**, 054515 (2007).
[19] D.C. Peets, J.D.F. Mottershead, B. Wu, I.S. Elfimov, R. Liang, W.N. Hardy, D.A. Bonn, M. Raudsepp, N.J.C. Ingle, A. Damascelli, New J. Phys. **9**, 28 (2007).
[20] J.G. Analytis, M. Abdel-Jawad, L. Balicas, M.M.J. French, N.E. Hussey, Phys. Rev. B **76**, 104523 (2007).
[21] J.M. Luttinger, J.C. Ward, Phys. Rev. **118**, 1417 (1960).
[22] C.M. Varma, Int. J. Mod. Phys. B **3**, 2083 (1989).
[23] J.L. Tallon, C. Bernhard, H. Shaked, R.L. Hitterman, J.D. Jorgensen, Phys. Rev. B **51**, 12911 (1995).
[24] A.F. Bangura, P.M.C. Rourke, T.M. Benseman, M. Matusiak, J.R. Cooper, N.E. Hussey, A. Carrington, Phys. Rev. B **82**, 140501 (2010).
[25] M. Platé, J.D.F. Mottershead, I.S. Elfimov, D.C. Peets, R. Liang, D.A. Bonn, W.N. Hardy, S. Chiuzbaian, M. Falub, M. Shi et al., Phys. Rev. Lett. **95**, 077001 (2005).
[26] A.A. Kordyuk, S.V. Borisenko, A. Koitzsch, J. Fink, M. Knupfer, B. Büchner, H. Berger, G. Margaritondo, C.T. Lin, B. Keimer et al., Phys. Rev. Lett. **92**, 257006 (2004).
[27] W.S. Lee, K. Tanaka, I.M. Vishik, D.H. Lu, R.G. Moore, H. Eisaki, A. Iyo, T.P. Devereaux, Z.X. Shen, Phys. Rev. Lett. **103**, 067003 (2009).
[28] M. Ossadnik, C. Honerkamp, T.M. Rice, M. Sigrist, Phys. Rev. Lett. **101**, 256405 (2008).
[29] P.A. Casey, P.W. Anderson, Phys. Rev. Lett. **106**, 097002 (2011).
[30] J.W. Loram, K.A. Mirza, J.M. Wade, J.R. Cooper, W.Y. Liang, Physica C **235-240**, 134 (1994).